# Shape-Based Magnetic Domain Wall Drift for an Artificial Spintronic Leaky Integrate-and-Fire Neuron

Wesley H. Brigner, *Student Member, IEEE*, Naimul Hassan, *Student Member, IEEE*, Lucian Jiang-Wei, Xuan Hu, *Student Member, IEEE*, Diptish Saha, *Student Member, IEEE*, Christopher H. Bennett, *Member, IEEE*, Matthew J. Marinella, *Senior Member, IEEE*, Jean Anne C. Incorvia, *Member, IEEE*, Felipe Garcia-Sanchez, *Member, IEEE*, and Joseph S. Friedman, *Member, IEEE*

*Abstract*—Spintronic devices based on domain wall (DW) motion through ferromagnetic nanowire tracks have received great interest as components of neuromorphic information processing systems. Previous proposals for spintronic artificial neurons required external stimuli to perform the leaking functionality, one of the three fundamental functions of a leaky integrate-and-fire (LIF) neuron. The use of this external magnetic field or electrical current stimulus results in either a decrease in energy efficiency or an increase in fabrication complexity. In this work, we modify the shape of previously demonstrated three-terminal magnetic tunnel junction neurons to perform the leaking operation without any external stimuli. The trapezoidal structure causes shape-based DW drift, thus intrinsically providing the leaking functionality with no hardware cost. This LIF neuron therefore promises to advance the development of spintronic neural network crossbar arrays.

*Index Terms*—Artificial neuron, leaky integrate-and-fire (LIF) neuron, magnetic domain wall, neural network crossbar, neuromorphic computing, three-terminal magnetic tunnel junction (3T-MTJ)

## I. INTRODUCTION

Modern von Neumann computing systems are capable of efficiently solving staggeringly difficult problems when provided with a structured data set. However, the human brain outperforms computers when processing unstructured real-world information. In fact, the brain is capable of performing these tasks with many orders of magnitude less energy than is required by computers [1]–[3]. This impressive computational efficiency is, according to neuroscientists, the result of complex interactions occurring between neurons and synapses. Neurons are complex nerve cells which integrate electrical signals received via the cells' dendrites, originate electrical signals (spikes) in the soma (cell body), and propagate these signals forward into their axons to convey information. Meanwhile, synapses are the electrically conductive junctions between the axon of one neuron and the dendrite of another and permit communication between neurons.

A primary objective in emulating neurobiological behavior within an artificial system is to efficiently replicate the neuron and synapse functionalities. This can be emulated with software running on standard computer hardware [4], [5], though such approaches consume significantly greater energy than their biological counterparts [6]. Energy improvements have therefore been demonstrated with dedicated hardware neural networks in which synapse and neuron functionalities are replicated with silicon transistors [2], [3]. However, the history-dependent nature of much synapse and neuron behavior inspire the use of non-volatile devices for increased efficiency. To that end, non-volatile devices such as memristors [7], magnetic skyrmion tracks [8], and three-terminal magnetic tunnel junctions (3T-MTJs) [9], [10] have been used that thoroughly mimic the functionalities of biological synapses. However, replicating the complex integrative and temporal behaviors occurring within a neuron's cell body (soma) has been a greater challenge.

The implementation of a single-device "leaky integrate-and-fire" (LIF) neuron has been inhibited by the need to provide the following functionalities:

1. Integration: accumulation of the input signal,
2. Firing: emission of a signal once the accumulated input signal reaches a particular threshold,
3. Leaking: continuous diminishing of the accumulated signal, and

Several artificial neurons have previously been proposed based on CMOS [11], magnetoelectric MTJs [12], spin-transfer torque random-access memory [13], and 3T-MTJs [10], [14]. While most prior work requires external circuitry to implement the leaking, we previously proposed an artificial neuron that provided all three behaviors with a single 3T-MTJ device [15].

This paper proposes an alternative 3T-MTJ neuron in which the fabrication is simplified by reducing the number of material layers: the leaking effect provided by the bottom ferromagnet is here provided instead by shape-based magnetic domain wall (DW) drift. This effect is explained in section III following neural network and 3T-MTJ background in section II; the neuron functionality is demonstrated in section IV, followed by conclusions in section V.

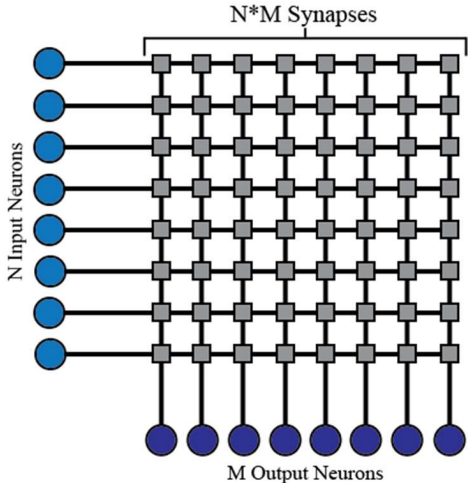

Fig. 1. 8x8 neural network. The eight input neurons and eight output neurons are connected via 64 synapses.

## II. BACKGROUND

A hardware neural network requires electrical analogs of biological neurons and synapses. These devices must be connected in such a fashion as to be compatible with standard fabrication processes, which gives rise to a synapse crossbar array structure that provides weighted connectivity between input and output neurons.

### A. Neural Network Crossbar Array

A crossbar neural network consists of synapses and LIF neurons, as shown in Fig. 1. These components are connected in an NxM crossbar array consisting of N horizontal wires (word lines) and M vertical wires (bit lines) such that the crossbar array contains N+M LIF neurons and N*M synapses. Neurons are placed at the inputs of the word lines and at the outputs of the bit lines, while the synapses are placed at the intersections between the word and bit lines. The individual states of the synapses determine the electrical connectivity between the various input and output neurons, and therefore the amount of current transmitted from the input neurons to the output neurons.

### B. Synapse

Synapses in a neural network express the degree of correlation or attraction between two neurons, and can be electrically modulated via conductivity. In the brain, a synapse, which is also electrically conductive, connects the gap between two neurons and controls communication between the two neurons it connects. The artificial analog of this biological component performs the same function – it bridges two neurons in a neural network. However, in this case the weight between these two neurons is determined during the training of the neural network. This behavior is readily provided by non-volatile devices with multiple resistance states, including memristors [16], [17] and 3T-MTJ artificial synapses [9], [10].

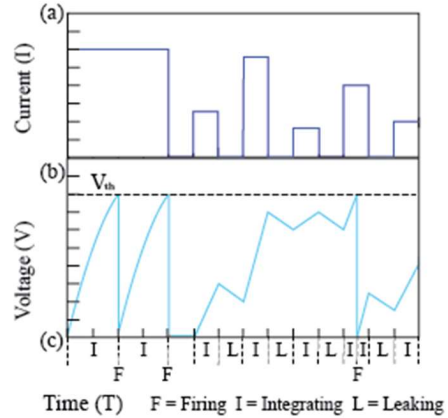

Fig. 2. Illustration of the functionality of an LIF neuron. An input current pulse train is shown in (a), while (b) shows the energy stored in the neuron by a voltage while leaking, integrating, and firing. Labels for the three states are provided in (c).

### C. Leaky Integrate-and-Fire Neuron

The LIF neuron has also received much interest as a representation of biological neuron behavior, and is a slight modification of the original integrate-and-fire neuron [15]. It emulates the behavior of a biological neuron by receiving and emitting current spikes. As the name implies, an LIF neuron exhibits three behaviors – integrating, leaking, and firing. In integration, a series of current spikes are applied to the input, causing the neuron's stored energy to increase. In leaking, which occurs whenever the input is inactive, the neuron gradually leaks a small amount of its stored energy. If presynaptic inputs have caused the neuron's energy to exceed a certain threshold level, it will then emanate a current spike, which flows outwards towards the various synapses connected to other neurons. The operation of the neuron is illustrated in Fig. 2, where the state of the neuron is represented by a voltage.

### D. Three-Terminal Magnetic Tunnel Junction

As shown in Fig. 3(a), The 3T-MTJ consists of a soft ferromagnetic nanowire track in which a DW can move relative to an MTJ [18], [19]. On either end of the DW track are regions of fixed magnetization that prevent the DW from being annihilated by either end of the nanowire. Additionally, an MTJ is formed by a tunnel barrier between the ferromagnetic nanowire and a fixed ferromagnet.

When a current flows through the nanowire track in the +$x$ (-$x$) direction, the DW travels in the opposite direction: -$x$ (+$x$). As a Néel DW moves, its magnetization state continually 'rotates', or precesses, out of the plane of the DW. For example, for the DW similar of Fig. 3, the DW to precess in the $x$-$y$ plane. Whenever the DW crosses under the tunnel barrier, the MTJ resistance across the tunnel barrier switches between the high-resistance anti-parallel state and low-resistance parallel state. While initially proposed for non-volatile Boolean logic, this device has more recently generated significant interest in the context of neuromorphic computing.

### E. 3T-MTJ Neuron with Intrinsic Leaking

Previously, we proposed an LIF neuron using a 3T-MTJ device with an additional ferromagnetic layer placed underneath the nanowire track [15]. The ferromagnet produces a constant magnetic field oriented in the $-z$-direction that causes the DW to gradually travel in the $-x$-direction. This allows the device to leak without the use of any additional current or control circuitry, enabling improved device and system efficiency. We also proposed a 3T-MTJ neuron using an anisotropy gradient [20]. This can be implemented using either Ga$^+$ ion irradiation or a TaO$_x$ wedge fabricated on top of the device.

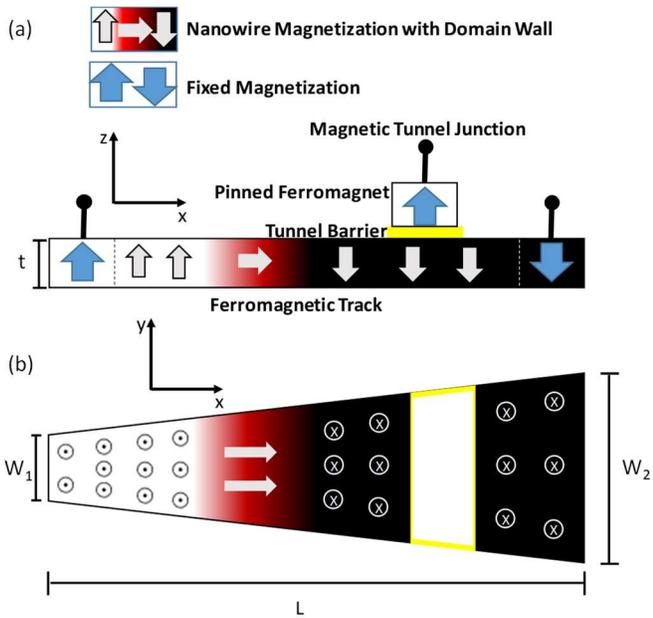

Fig. 3. Side (a) and top (b) views of the proposed device. The side view is identical to that of a conventional 3T-MTJ.

## III. Intrinsically Leaking 3T-MTJ Device with Shape-Based DW Drift

DWs in conventional 3T-MTJ devices have previously been moved only with externally-applied currents or magnetic fields. However, by modifying the shape of a standard 3T-MTJ device, the DW can be made to move autonomously as a result of the non-uniform DW energy landscape resulting from the non-uniform device width.

### A. Device Structure

We propose a novel spintronic neuron that is identical to previously-described 3T-MTJ devices [18], [19] in all but one respect: instead of a rectangular $x$-$y$ cross-section, the ferromagnetic track has a trapezoidal $x$-$y$ cross-section. While the side view of Fig. 3(a) is unchanged, the top view of Fig. 3(b) makes this modification clear. Similar to the conventional 3T-MTJ, there is a Néel DW, in which the magnetization is in the $-z$-direction to the left of the DW and the $+z$-direction to the right of the DW.

The micromagnetic simulations described in this paper were performed with Mumax3 [21], with length $L$ of 250 nm, left-hand width $W_1$ of 25 nm, right-hand width $W_2$ of 100 nm, and thickness t of 1.5 nm. The fixed magnetizations at either end of the ferromagnetic nanowire cover 10 nm from each edge, providing the DW with a 230 nm range of motion. The material parameters represent CoFeB, with an exchange stiffness $A_{ex}$ of $13\times10^{-12}$ J/m, a Landau-Lifshitz-Gilbert damping constant $\alpha$ of 0.05, a non-adiabaticity factor $\xi$ of 0.05, a magnetic saturation value $M_{sat}$ of 1 T, and a uniaxial anisotropy in the z-direction with a magnitude of $5\times10^5$ J/m$^3$. The cell size is 1x1x1.5 nm$^3$, and the external magnetic field $B_{ext}$ is 0 T everywhere. The COMSOL multiphysics simulator was used to determine the electrical current density through this trapezoidal structure.

### B. Leaking with Shape-Based DW Drift

Because of the trapezoidal structure of the ferromagnetic nanowire, the energy of a DW is dependent on the position of the DW along the length of the track [22]. In particular, the DW energy depends on the shape anisotropy of the magnetic material, and the asymmetric shape modifies the demagnetization factor of the magnetic structure. The DW energy is highest where the width is largest, and is lowest where the width is smallest. Therefore, in order to minimize the DW energy, the DW autonomously moves leftward from higher-energy positions at the right (wide) side of the wire to lower-energy positions at the left (narrow) side of the wire.

This DW motion is observed in the micromagnetic simulation of Fig. 4. The DW is initialized 175 nm from the narrow left edge of the nanowire track (75 nm from the wide right edge), and gradually drifts towards the narrow left edge. No electrical current or magnetic field is applied. The DW precesses as it drifts, generating the ripple seen in the position over time; the DW maintains a steady-state position while continually precessing once it reaches the stable position 28 nm from the left edge. It can further be seen that the DW velocity increases as the DW approaches the narrow edge of the nanowire.

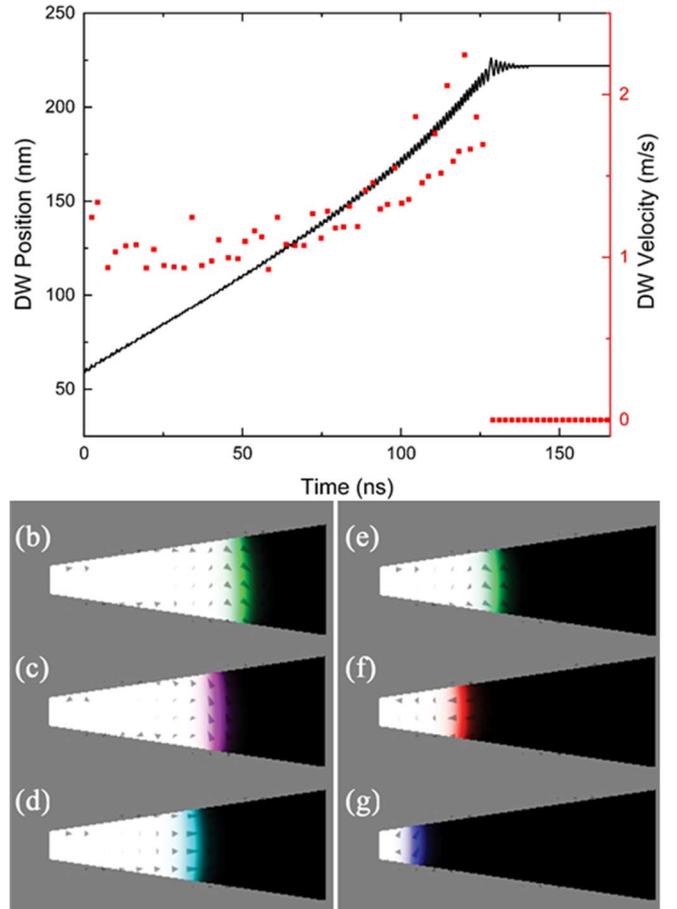

Fig. 4. Shape-based DW drift with no external stimuli. (a) Position and instantaneous velocity of the DW as functions of time. Inset: Velocity as a function of the DW width. The position was calculated based on the minimum of the absolute value of the z-directed magnetization along the central axis of the nanowire length; the velocity was determined using an average DW position calculated from the local maxima and minima caused by the precession of the DW. Micromagnetic simulation snapshots are shown for (b) t = 0, (c) t = 22 ns, (d) t = 44 ns, (e) t = 66 ns, (f) t = 88 ns, and (g) t = 110 ns.

### C. Experimental Considerations

The field-free and current-free movement of the DW from a wider to narrower region of the ferromagnetic track depends on the energy difference of the demagnetization field due to the asymmetric shape compared to the pinning energy of the DW due to intrinsic and extrinsic defects in the wire, for example from dopants and edge roughness [23]. The simulations described throughout this paper were performed at zero temperature in a perfect wire without these pinning effects. Experimental demonstration of the proposed neuron at room temperature [24] should therefore be feasible with a sufficiently pristine nanowire.

## IV. ARTIFICIAL NEURON WITH SHAPE-BASED DW DRIFT

The shape-based DW drift provides a native representation of neuron leaking that enables simplification of the device structure. Whereas previous spintronic neuron proposals have required external currents, magnetic fields, or additional device layers, the shape-based DW drift enables an artificial 3T-MTJ neuron with an intrinsic leaking capability. The integration and firing capabilities are retained in a manner similar to previous proposals [14], rounding out the requirements for an LIF neuron.

### A. Integration of Externally-Applied Current

As described in [15], current through the device is integrated through motion of the DW. The DW velocity is dependent on the applied current as shown in Fig. 5, with larger currents causing faster integration of the externally-applied signal. With this trapezoidal prism, the DW velocity is also influenced by the width, as discussed previously in relation to the leaking; the DW moves faster where the width is smaller.

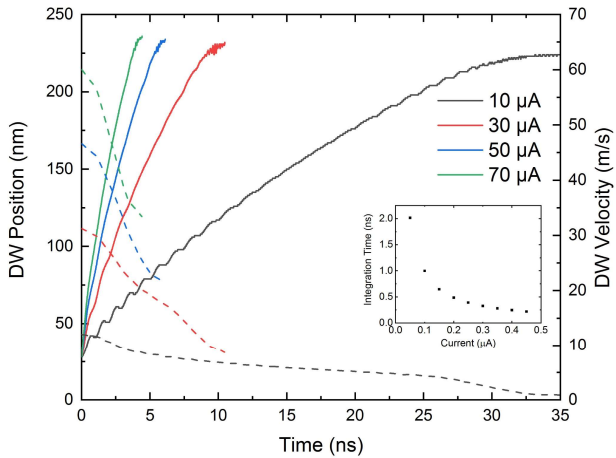

Fig. 5. (a) The position and instantaneous velocity of the DW for various currents, calculated similarly to Fig. 4. Positions are represented as solid curves, while velocities are represented by dashed curves. Inset: The time taken for a DW to shift 100 nm from the stable (28 nm from the left edge of the device) as a function of the current passed through the DW track.

### B. Integration & Leaking with Shape-Based DW Drift

The combined integration and leaking behavior of the proposed neuron is demonstrated in Fig. 6. A 2 ns period of integration with a 50 µA current is followed by a 30 ns period of leaking during which no current flows through the ferromagnetic track. This pattern repeats twice for a total runtime of 96 ns. As can be seen in the simulation results, the DW position increases rapidly when current is applied during the integration periods, and precesses while decreasing gradually when leaking in the absence of any external stimuli.

### C. Firing through Magnetoresistance Switching

In an LIF neuron, the firing commences when enough energy has been stored in the neuron. In the case of the proposed 3T-MTJ neuron, firing occurs when the DW has passed underneath the tunnel barrier and fixed ferromagnet, switching the MTJ from its high-resistance state to its low-resistance state. This state change can provide a voltage pulse that can be used as an output spike that provides a current pulse to downstream synapses and neurons.

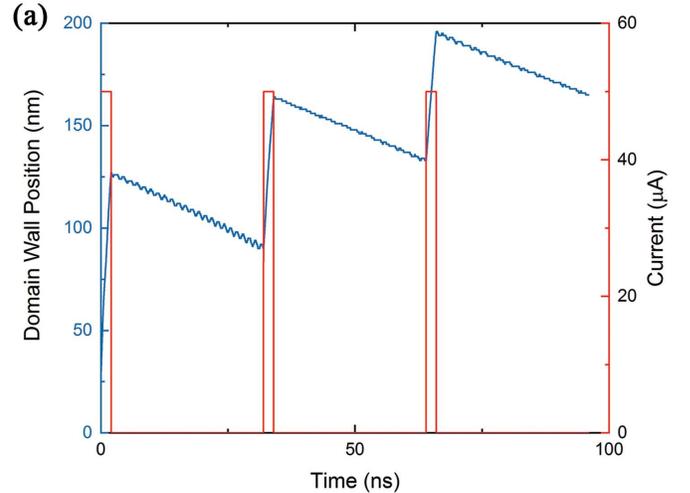

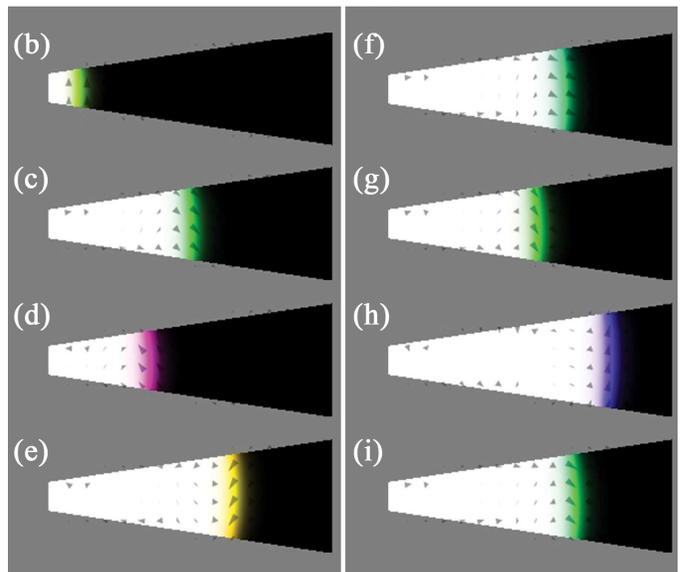

Fig. 6. (a) Applied current and DW position as a function of time, demonstrating the leaking and integrating functionalities of the neuron. (b) Micromagnetic simulation snapshots are shown for (b) t = 0, (c) t = 2 ns, (d) t = 17 ns, (e) t = 32 ns, (f) t = 34 ns, (g) t = 64 ns, (h) t = 66 ns, and (i) t = 96 ns.

## V. Conclusion

This work proposes an LIF neuron that uses shape-based magnetic DW drift to leak without any external stimuli by altering the shape of a standard 3T-MTJ device. The proposed trapezoidal structure creates regions of higher and lower energy states for the DW, causing the DW to autonomously drift toward the narrower edge that corresponds to lower DW energies. This proposed neuron is an improvement over previous spintronic LIF neurons, which require an additional ferromagnetic layer or the application of electrical current to induce the neuron leaking. This structure provides improvements in both power dissipation and ease-of-fabrication, and is therefore an important potential building block of future neuromorphic information processing systems.

## Acknowledgments

The authors acknowledge technical support from E. Laws, J. Mcconnell, N. Nazir, and L. Philoon.

This paper describes objective technical results and analysis. Any subjective views or opinions that might be expressed in the paper do not necessarily represent the views of the U.S. Department of Energy or the United States Government.

## References

[1] V. Balasubramanian, "Heterogeneity and efficiency in the brain," *Proc. IEEE*, vol. 103, no. 8, pp. 1346–1358, Aug. 2015.

[2] F. Akopyan *et al.*, "TrueNorth: Design and tool flow of a 65 mw 1 million neuron programmable neurosynaptic chip," *IEEE Trans. Comput. Des. Integr. Circuits Syst.*, vol. 34, no. 10, pp. 1537–1557, Oct. 2015.

[3] P. A. Merolla *et al.*, "A million spiking-neuron integrated circuit with a scalable communication network and interface," *Science*, vol. 345, no. 6197, pp. 668–673, 2014.

[4] A. Delorme, J. Gautrais, R. van Rullen, and S. Thorpe, "SpikeNET: A simulator for modeling large networks of integrate and fire neurons," *Neurocomputing*, vol. 26–27, pp. 989–996, Jun. 1999.

[5] S. Han, H. Mao, and W. J. Dally, "Deep compression: compressing deep neural networks with pruning, trained quantization and huffman coding," 2016.

[6] B. Sengupta and M. B. Stemmler, "Power consumption during neuronal computation," *Proc. IEEE*, vol. 102, no. 5, pp. 738–750, May 2014.

[7] D. B. Strukov, G. S. Snider, D. R. Stewart, and R. S. Williams, "The missing memristor found," *Nature*, vol. 453, pp. 80–83, May 2008.

[8] X. Chen, W. Kang, D. Zhu, X. Zhang, N. Lei, Y. Zhang, Y. Zhou, and W. Zhao, "A compact skyrmionic leaky–integrate–fire spiking neuron device," *Nanoscale*, vol. 10, no. 13, pp. 6139–6146, 2018.

[9] S. Dutta, S. A. Siddiqui, F. Buttner, L. Liu, C. A. Ross, and M. A. Baldo, "A logic-in-memory design with 3-terminal magnetic tunnel junction function evaluators for convolutional neural networks," in *2017 IEEE/ACM International Symposium on Nanoscale Architectures (NANOARCH)*, 2017, pp. 83–88.

[10] A. Sengupta, Y. Shim, and K. Roy, "Proposal for an all-spin artificial neural network: Emulating neural and synaptic functionalities through domain wall motion in ferromagnets," *IEEE Trans. Biomed. Circuits Syst.*, vol. 10, no. 6, pp. 1152–1160, 2016.

[11] D. Querlioz, W. S. Zhao, P. Dollfus, J.-O. Klein, O. Bichler, and C. Gamrat, "Bioinspired networks with nanoscale memristive devices that combine the unsupervised and supervised learning approaches," in *Proceedings of the 2012 IEEE/ACM International Symposium on Nanoscale Architectures - NANOARCH '12*, 2012, pp. 203–210.

[12] A. Jaiswal, S. Roy, G. Srinivasan, and K. Roy, "Proposal for a leaky-integrate-fire spiking neuron based on magnetoelectric switching of ferromagnets," *IEEE Trans. Electron Devices*, vol. 64, no. 4, pp. 1818–1824, Apr. 2017.

[13] A. Sengupta and K. Roy, "Spin-transfer torque magnetic neuron for low power neuromorphic computing," in *2015 International Joint Conference on Neural Networks (IJCNN)*, 2015, vol. 2015–Septe, pp. 1–7.

[14] A. Sengupta and K. Roy, "A vision for all-spin neural networks: a device to system perspective," *IEEE Trans. Circuits Syst. I Regul. Pap.*, vol. 63, no. 12, pp. 2267–2277, Dec. 2016.

[15] N. Hassan *et al.*, "Magnetic domain wall neuron with lateral inhibition," *J. Appl. Phys.*, vol. 124, no. 15, p. 152127, Oct. 2018.

[16] Y.-P. Lin *et al.*, "Physical realization of a supervised learning system built with organic memristive synapses," *Sci. Rep.*, vol. 6, no. 1, p. 31932, Oct. 2016.

[17] M. Sharad, D. Fan, K. Aitken, and K. Roy, "Energy-efficient non-boolean computing with spin neurons and resistive memory," *IEEE Trans. Nanotechnol.*, vol. 13, no. 1, pp. 23–34, Jan. 2014.

[18] J. A. Currivan-Incorvia *et al.*, "Logic circuit prototypes for three-terminal magnetic tunnel junctions with mobile domain walls," *Nat. Commun.*, vol. 7, p. 10275, Jan. 2016.

[19] J. A. Currivan, Youngman Jang, M. D. Mascaro, M. A. Baldo, and C. A. Ross, "Low energy magnetic domain wall logic in short, narrow, ferromagnetic wires," *IEEE Magn. Lett.*, vol. 3, pp. 3000104–3000104, 2012.

[20] W. H. Brigner, X. Hu, N. Hassan, C. H. Bennett, J. A. C. Incorvia, F. Garcia-Sanchez, and J. S. Friedman, "Graded-Anisotropy-Induced Magnetic Domain Wall Drift for an Artificial Spintronic Leaky Integrate-and-Fire Neuron," *IEEE Journal on Exploratory Solid-State Computational Devices and Circuits*, pp. 1–1, 2019.

[21] A. Vansteenkiste, J. Leliaert, M. Dvornik, M. Helsen, F. Garcia-Sanchez, and B. Van Waeyenberge, "The design and verification of MuMax3," *AIP Adv.*, vol. 4, no. 10, p. 107133, Oct. 2014.

[22] J. Wunderlich *et al.*, "Influence of geometry on domain wall propagation in a mesoscopic wire," *IEEE Trans. Magn.*, vol. 37, no. 4, pp. 2104–2107, Jul. 2001.

[23] S. Dutta, S. A. Siddiqui, J. A. Currivan-Incorvia, C. A. Ross, and M. A. Baldo, "Micromagnetic modeling of domain wall motion in sub-100-nm-wide wires with individual and periodic edge defects," *AIP Adv.*, vol. 5, no. 12, 2015.

[24] D. Vodenicarevic *et al.*, "Low-energy truly random number generation with superparamagnetic tunnel junctions for unconventional computing," *Phys. Rev. Appl.*, vol. 8, no. 5, p. 054045, Nov. 2017.